%% file: main_wcnc_v2_arxiv.tex
\documentclass[conference]{IEEEtran}
\usepackage{graphicx, epsfig,amsmath,amssymb,latexsym,graphics,float}
\usepackage{cite}
\usepackage{setspace,subfig}
\usepackage{lipsum}
\usepackage{multicol}
\usepackage{bbding}
\begin{document}
\begin{onecolumn}
\input{title-track}
\end{onecolumn}
\newpage
\begin{twocolumn}
\title{Online Fault-Tolerant Dynamic Event Region Detection in Sensor Networks via Trust Model}
\author{\IEEEauthorblockN{Jiejie Wang}
\IEEEauthorblockA{School of Computer Science and Technology\\
Nanjing University of Posts and Telecommunications\\
Nanjing, Jiangsu 210023, China}
\and
\IEEEauthorblockN{Bin Liu$^($\Envelope $^)$}
\IEEEauthorblockA{School of Computer Science and Technology\\
Nanjing University of Posts and Telecommunications\\
Nanjing, Jiangsu 210023, China\\
Email: bins@ieee.org}}
\maketitle
\begin{abstract}
This paper proposes a Bayesian modeling approach to address the problem of online fault-tolerant dynamic event region detection in wireless sensor networks. In our model every network node is associated with a virtual community and a \emph{trust} index, which quantitatively measures the trustworthiness of this node in its community. If a sensor node's \emph{trust} value is smaller than a threshold, it suggests that this node encounters a fault and thus its sensor reading can not be trusted at this moment. This concept of sensor node \emph{trust} discriminates our model with the other alternatives, e.g., the Markov random fields. The practical issues, including spatiotemporal correlations of neighbor nodes' sensor readings, the presence of sensor faults and the requirement of online processing are linked together by the concept \emph{trust} and are all taken into account in the modeling stage.
Based on the proposed model, the \emph{trust} value of each node is updated online by a particle filter algorithm upon the arrival of new observations. The decision on whether a node is located in the event region is made based upon the current estimate of this node's \emph{trust} value. Experimental results demonstrate that the proposed solution can provide striking better performance than existent methods in terms of error rate in detecting the event region.
\end{abstract}
\IEEEpeerreviewmaketitle
\section{Introduction}
Event region detection (ERD) is a key problem for many real applications of wireless sensor networks (WSNs), such as environment monitoring \cite{werner2005monitoring}, battlefield surveillance \cite{dhillon2003sensor}, structural health monitoring \cite{kim2006wireless} and so on.
An event can be anything that reflects an improper situation according to collective sensory data with respect to certain criteria.
An early enough detection of an unusual event as well as an accurate localization of the event region could be crucial for precluding the appearance of catastrophic events.

ERD has been explored relatively extensively in the static setting, in which the event region is assumed to be time invariant \cite{krishnamachari2004distributed,chen2005comments,wang2008collaborative,luo2006distributed,ermis2005adaptive,ermis2006detection,yim2010adaptive}, while, this assumption is not realistic. For instance, for events like forest fire, leakage of poisonous gas or debris flow, in fact the event region changes over time. To take into account of the dynamic nature of real events, several efforts have been made recently \cite{wu2011distributed,wu2014online,wu2014adaptive,Wu_2013_thesis}. The basic idea is to employ the information of the system dynamics, usually modeled by markov random fields (MRF) \cite{wang2005collaborative,yin2009spatio}, together with information collected from neighbor nodes to predict the
underlying hypothesis at each node, and then use its local observation for update.

For dynamic ERD problems, a fault-tolerant solution is always preferable, since many of real-world deployments of WSNs experience high rates of failure \cite{enemark2015energy,luo2006distributed,yao2007temporal}. Despite of many efforts that have been made in the literature, it is still a big challenge to do fault-tolerant, online, distributed and dynamic ERD for WSNs. To this end, we propose a novel model based approach to address the above challenge. A concept termed \emph{trust} plays a key role here and discriminates our model with the other alternatives in the literature.
Inspired by the concept of social \emph{trust}, which qualitatively describes the relationships among the members of a community based upon their past interactions, the concept \emph{trust} is employed here as a quantitative dynamic parameter to measure the trustworthiness of WSN nodes.
To the best of our knowledge, this is the first application of a trust model in the problem of dynamic ERD in WSNs. A Bayesian filtering procedure is performed to estimate the \emph{trust} online based on observations of a local community. The decision on whether a node is located within the event region is made based upon the \emph{trust} of this node.
With aid of the introduced \emph{trust} concept, the spatiotemporal correlations in neighbor nodes' sensor readings, sensor faults, the requirement of online processing are all linked together and taken into account in the modeling stage. Further, the proposed solution distributes computations to each sensor node and thus has a desirable scaling property, making it be able to handle large scale networks.
\section{Problem description}\label{sec:problem}
In the WSN under our consideration, each network node is comprised of a sensor, a wireless communication module to connect with close-by nodes, a processing unit and some storage.
Each node is associated with a virtual community of network nodes spatially centered around it. In what follows, we term the node of our interest in a community as center node, and the others are named member nodes. The member nodes are spatial neighbors of the center node, and they all fall within the communication range of the center node. Perfect data transmission between the center node and its corresponding member nodes is assumed here.  The occurrence of possible events corresponds to a set of neighboring network nodes whose sensory data deviate from a normal sensing range in a collective fashion.

In our model, which will be detailed in Sec.\ref{sec:model}, the center node is associated with a \emph{trust} index. If its \emph{trust} value is smaller than a threshold (empirically set at 0.3 in our experiments), then it suggests that the reading of the center node does not conform to some attribute embodied by readings of the member nodes. A probability density function (pdf) is used to model the uncertainty in the estimate of the \emph{trust} value. A Bayesian filtering procedure, which will be detailed in Sec.\ref{sec:alg}, is performed to update this pdf upon the arrival of new observations.
At each time step, the center node sends its node ID along with a one-bit encoded binary event status indicator to a close-by relay node, which may be a member node of its community. The relay node then transmits the received data to sink node(s). When all event indicator data are collected at sink node(s), then the critical event region is detected and reconstructed. Upon the arrival of new data at sink node(s), the event region will be updated.

To make the above workflow run, two key problems need to be resolved, namely, how to model the \emph{trust} with an effort to make full use of the related prior knowledge; and, given the model, how to estimate the \emph{trust} timely, accurately and efficiently upon the arrival of new observations. We address the above problems in detail in Sec. \ref{sec:model} and Sec. \ref{sec:alg}, respectively.
\section{Bayesian Trust Model (BTM)}\label{sec:model}
We adapt the model presented in \cite{liu2015toward} to deal with the problem of dynamic ERD in WSNs.
We focus on a single community of network nodes.
Following \cite{liu2015toward}, we use $\lambda$, $0\leq\lambda\leq1$ to denote the \emph{trust} of the center node. This variable quantitatively measures the extent to which the center node is trusted by a virtual third party that is assumed to be completely fair.
When $\lambda=1$, it suggests that the reading of the center node is completely reliable, while $\lambda=0$ indicates the opposite. Denote $D_k\triangleq\{x_{k,0}, x_{k,1},x_{k,2}\ldots,x_{k,n_k}\}$ to be the sensor readings collected by the center node at time step $k$, where $x_{k,0}$ denotes the reading of the center node itself, $x_{k,i}, i>0$ the reading of its $i$th community member and $n_k$ the number of its community members, at time step $k$.

We model the time-evolution law of $\lambda$ by
\begin{equation}\label{transition}
\lambda_{k}=\alpha\lambda_{k-1}+v,\quad v\sim\mathcal{N}(0,Q),
\end{equation}
where $v$ denotes a truncated zero-mean Gaussian noise with variance $Q$ and $0\leq\alpha\leq1$ is
a free parameter used to control the rate at which an old \emph{trust} should be \emph{forgotten}. An empirically choice of the $\alpha$ value is 0.85 as presented in \cite{liu2015toward}. The truncated Gaussian noise model is used here to guarantee that the value of $\lambda_{k}$ falls within the range [0,1]. The relationship between $\lambda_{k}$ and $D_k$ is formulated by a likelihood function as follows
\begin{equation}\label{likelihood}
p(D_k|\lambda_{k})=\exp\left(\frac{-\mid\lambda_{k}-V_{k}\mid}{\beta}\right)
\end{equation}
where $0<\beta<1$ is a preset model parameter and $V_{k}$ is defined to be
\begin{equation}\label{eqn:voting}
V_{k}\triangleq \frac{\sum_{i=1}^{n_k} U(i,D_k)}{n_k},
\end{equation}
in which
\begin{equation}\label{eqn:u}
U(i,D_k)=\left\{\begin{array}{ll}
1,\quad\mbox{if}\quad|x_{k,i}-x_{k,0}|< r \\
0,\quad\mbox{otherwise} \end{array} \right.
\end{equation}
where $r$ denotes a preset constant, representing the maximum permissible
difference in readings of a pair of mutually trusted nodes within a community.
If $U(i,D_k)$ equals 1, then it represents that the $i$th community member casts a vote of that the center node is trusted at time $k$. The
definition of $U(i,D_k)$ is inspired by the concept of social \emph{trust}, which tells that a pair of mutually trusted social entities within a community should have similar opinions on a characteristic object or event associated with this community. Such a definition of $U(i,D_k)$ can also be understood as an implicit way of employing the prior knowledge on the spatial correlations among sensor readings generated from the same community. The parameter $V_{k}$ in Eqn. (\ref{eqn:voting}) is defined to be an average of the voting results casted by those member nodes.
Given a specific value of $\lambda_{k}$, the exponential function adopted in Eqn.(\ref{likelihood}) renders the likelihood of $\lambda_{k}$ highly sensitive to the value of $V_{k}$. The closer the distance between $\lambda_{k}$ and $V_{k}$, the bigger is $\lambda_{k}$'s likelihood. The degree of sensitivity is controlled by the parameter $\beta$.

Now we treat the problem of online \emph{trust} estimation from the perspective of Bayesian state filtering, which consists of calculating the \emph{a posteriori} pdf of $\lambda_{k}$ given $D_{1:k}\triangleq\{D_1,\ldots,D_k\}$, denoted by $p(\lambda_k|D_{1:k})$ (or in short $p_{k|k}$). Recursive solutions are desirable for online state filtering problems, and, indeed, $p_{k|k}$ can be computed from $p_{k-1|k-1}$ recursively as follows \cite{doucet2000sequential}
\begin{equation}\label{eqn:filter}
p_{k|k}=\frac{p(D_k|\lambda_k)\int p(\lambda_k|\lambda_{k-1})p_{k-1|k-1}d\lambda_{k-1}}{p(D_k|D_{1:k-1})},
\end{equation}
where $p(\lambda_k|\lambda_{k-1})$ denotes the state transition prior, which is determined by Eqn. (\ref{transition}).
According to the Bayesian philosophy, the posterior pdf $p_{k|k}$ consists of all the related information about $\lambda_k$ at the time step $k$ \cite{gelman2014bayesian}. So the remaining task is to derive $p_{k|k}$ from $p_{k-1|k-1}$, supposing that $p_{k-1|k-1}$ is \emph{a priori} known at time step $k$.
In the next Section, the particle filtering algorithm is presented as a generic approach to simulate $p_{k|k}$, given a particle approximation of $p_{k-1|k-1}$.
\section{BTM based Dynamic ERD Algorithm}\label{sec:alg}
\subsection{Algorithm Description}
The proposed BTM based algorithm is carried out in real time at each single network node. The output of the algorithm at each time step, say $k$, is a binary event indicator denoted by $e_k$, $e_k\in\{0,1\}$. If $e_k=1$, it means that the associated node is located in the event region, and vice versa. At each time step, the binary event indicators and their corresponding node IDs will be transmitted to sink node(s) via the relay nodes. Then the event region shall be reconstructed at the sink node(s).

Now let us focus on a single node and the community centered at it. In what follows, an algorithm based on particle filter and the model presented in Sec. \ref{sec:model} is presented. To begin with, let $\{\lambda_{k}^i, w_{k}^i\}_{i=1}^N$ denotes a random measure that characterises the posterior pdf $p_{k|k}$, which means
\begin{equation}
p_{k|k}\approx\sum_{i=1}^Nw_{k}^i\delta(\lambda_{k}-\lambda_{k}^i),
\end{equation}
where $\delta(\cdot)$ denotes the Dirac delta function, $\{\lambda_{k}^i\}_{i=1}^N$ is the set of support particles with associated weights $\{w_{k}^i\}_{i=1}^N$ satisfying $w_{k}^i>0, \forall i$ and $\sum_{i=1}^N w_{k}^i=1$. $N$ is the particle size.

Now let us consider the sequential case of our concern. Assume that at time $k-1$, a discrete weighted approximation to $p_{k-1|k-1}$, namely $\{\lambda_{k-1}^i, w_{k-1}^i\}_{i=1}^N$, is available, the question is how to derive a updated particle set to approximate $p_{k|k}$.
We resort to sequential importance sampling \cite{doucet2000sequential}, and select $p(\lambda_{k}|\lambda_{k-1})$ as the proposal distribution to draw new particles. Specifically, given $\lambda_{k-1}^i$, $\lambda_{k}^i$ is generated according to the time evolution law given by Eqn.(\ref{transition}) as follows
\begin{equation}
\lambda_{k}^i=\alpha\lambda_{k-1}^i+v^i, \quad\mbox{for}\quad i=1,\ldots,N,
\end{equation}
where $v^i$ is a random sample drawn from $\mathcal{N}(0,Q)$. If the value of $\lambda_{k}^i$ does not fall within the value space $[0,1]$, then a new particle will be drawn until it satisfies the above restriction. According to the theorem of importance sampling \cite{arulampalam2002tutorial}, the corresponding importance weight with respect to $\lambda_{k}^i$ is
\begin{equation}\label{eqn:is_weight}
w_{k}^i=\frac{p(D_k|\lambda_{k}^i)}{\sum_{j=1}^Np(D_k|\lambda_{k}^j)}, i=1,2,\ldots,N,
\end{equation}
then the resulting particle set $\{\lambda_{k}^i, w_{k}^i\}_{i=1}^N$ can provide a discrete weighted approximation to $p_{k|k}$. A resampling procedure can be optionally inserted here in order to avoid particle divergence \cite{arulampalam2002tutorial}.
Note that there are other ways to generate new particles and calculate the corresponding weights in the PF framework, see details in e.g., \cite{arulampalam2002tutorial,doucet2009tutorial,van2000unscented,liu2008particle}, while the algorithm presented here requires minimum costs in computing and storage, and thus is more suitable for WSN applications.

Given $\{\lambda_{k}^i, w_{k}^i\}_{i=1}^N$, we calculate the mean of the particles
\begin{equation}\label{eqn:lambda}
\hat{\lambda}_k=\sum_{i=1}^Nw_{k}^i\lambda_{k}^i,
\end{equation}
 which is the \emph{a posterior} and minimum mean squared error estimate of the true value of $\lambda_{k}$.
 Then we compare $\hat{\lambda}_k$ with a prescribed threshold $\lambda_{thr}$. If $\hat{\lambda}_k>\lambda_{thr}$, then we determine that the sensor reading of the center node, i.e., $x_{k,0}$, is trusted; otherwise, it is non-trusted. Assume that the normal sensing range is \emph{a priori} known as $[x_{\min},x_{\max}]$, the value of the event indicator, $e_k$, is determined as follows
\begin{equation}\label{eqn:e}
e_k=\left\{\begin{array}{ll}
0,\quad\mbox{if}\quad\hat{\lambda}_k>\lambda_{thr}, x_{k,0}\in [x_{\min},x_{\max}];\\
1,\quad\mbox{else if}\quad\hat{\lambda}_k>\lambda_{thr}, x_{k,0}\notin [x_{\min},x_{\max}];\\
0,\quad\mbox{else if}\quad \hat{x}_{k,0}\in [x_{\min},x_{\max}];\\
1,\quad\mbox{otherwise} \end{array} \right.
\end{equation}
in which
\begin{equation}\label{eqn:estimate_x_k}
\hat{x}_{k,0}=\frac{\sum_{j=1}^{n_k}\hat{\lambda}_{k,j}x_{k,j}}{\sum_{j=1}^{n_k}\hat{\lambda}_{k,j}}
\end{equation}
where $\hat{\lambda}_{k,j}$ denotes the \emph{trust} value of the $j$th member node at time step $k$. Note that $\hat{\lambda}_{k,j}$ is calculated out when this $j$th member node of this community acts as the center node of another community centered on it.
The value assignment rules embedded in Eqn.(\ref{eqn:e}) are delineated as follows. The first two rules specified in Eqn. (\ref{eqn:e}) correspond to cases associated with $\hat{\lambda}_k>\lambda_{thr}$, which suggests that the sensor reading of the center node, $x_{k,0}$, is likely to be trusted. In such cases, the value of $e_k$ to be assigned is totally dependent on whether the value of $x_{k,0}$ falls within the normal sensing range, $[x_{\min},x_{\max}]$. These two rules are commensurate with our intuition and easy to understand. The last two rules of Eqn.(\ref{eqn:e}) correspond to cases associated with $\hat{\lambda}_k\leq\lambda_{thr}$, which indicates that $x_{k,0}$ is likely to be generated by a sensor fault and thus is not trusted. For this scenario, we resort to the member nodes to generate a trusted estimate of $x_{k,0}$. Specifically we calculate a \emph{trust}-averaged sensor reading, $\hat{x}_{k,0}$, by Eqn.(\ref{eqn:estimate_x_k}), and take it as an estimate of a trusted sensor reading that should be. Then we determine the value of $e_k$ based on the fact whether the value of $\hat{x}_{k,0}$ falls within the normal sensing range.

The above operations are associated with one specific network node. When the same operations are completed on each node of the network, then we get the $e_k$s of all the network nodes. These $e_k$s are then transmitted to the sink node, wherein the event region will be updated accordingly.
\subsection{Theoretical Analysis}
The purpose of the analysis is to endow readers with a conceptual understanding of the way in which the proposed algorithm disambiguates fault-event in the context of dynamic ERD. Here fault is defined to be a phenomenon that suddenly happens at a network node, rendering the sensor reading of this node not commensurate with the spatiotemporal regularity held within its community. Different from faulty data, normal sensor readings associated with either the event or the non-event region are assumed to have specific spatiotemporal regularities.

First, let us analyze how the proposed algorithm handles the node faults that appear in the non-event region. Imagine a network node within the non-event region that has been working normally before time $k$; hence the support of its \emph{a prior} pdf of the \emph{trust}, $p(\lambda_k|\lambda_{k-1})$, tends to the maximum \emph{trust} value at 1. Assume that this node suffers a sudden failure and its sensor reading deviates from the normal sensing range at time step $k$. As a result, the $V_{k}$ defined in Eqn.(\ref{eqn:voting}) will get a relatively small value. Then the definition of the likelihood (see Eqn. (\ref{likelihood})) determines that only very small valued $\lambda_k^i$s can obtain importance weights that are relatively larger than 0, because the importance weight is proportional to the likelihood as presented in Eqn. (\ref{eqn:is_weight}). So the value of $\hat{\lambda}_k$, determined by Eqn. (\ref{eqn:lambda}), is likely to get a very small value and finally this node will be taken as a fault node. According to the third rule specified in Eqn. (\ref{eqn:e}), $e_k$ will be set at 0, indicating that this node is located in the non-event region.

Next let us consider the situation in which the center node is located in the event region and then encounters a sudden fault at time step $k$. Since the faulty measurement is not consistent with the regularities held by sensor readings of the majority of member nodes, which are operating normally, $V_{k}$ of the center node, defined in Eqn. (\ref{eqn:voting}), shall get a relatively small value. Performing a similar analysis as before, we can predict that the algorithm will recognize this node to be a faulty node. Finally, according to the last rule specified in Eqn. (\ref{eqn:e}), $e_k$ will be set at 1, indicating that this node is located in the event region.

Finally, let us take a look at the working mechanism of the proposed algorithm at the boundary nodes.
As the event region is allowed to dynamically change over time, there exist two types of boundary nodes, namely those that have just entered the event region and that have just left. For each boundary node, some of its corresponding member nodes are located in the event region and the others are not. Assume that the spatial distribution of the network nodes is uniform over the monitoring space. The voting result provided by the member nodes is probable to be neutral. Then the posterior pdf of $\lambda_k$ will be largely dependent on the prior. So provided that this node worked normally before and obtain a good reputation, the posterior estimate of its \emph{trust}, $\hat{\lambda}_k$, will be bigger than $\lambda_{thr}$ (which is empirically set at 0.3 in our experiments). As a result, the value of $e_k$ will be only dependent on the value of $x_{k,0}$, according to the first two rules specified in Eqn. (\ref{eqn:e}).
Therefore, provided that this node still works normally at this moment, the algorithm will output a correct $e_k$ value. If the sensor failure happens to occur at this boundary node, it is probable to result in a false alarm or miss-detection. However, the above phenomenon can have only a limited impact on the overall performance, because the number of boundary nodes is always much smaller than that of the whole network nodes.

The above theoretical analysis was corroborated by experimental results in simulation studies, which are presented in the next section.
\section{Performance Evaluation by Simulations}\label{sec:simu}
To evaluate the performance of the proposed algorithm, we
performed experiments on synthetic data generated by simulations.
\subsection{Baselines and Metrics}
We use two different algorithms as the baselines for comparison. The first baseline is the traditional cumulative sum (CUSUM) detector, a widely used algorithm for abrupt change detection, which has also been studied extensively in the context of WSNs \cite{ying2014cusum,de2010cusum,wu2011distributed}. The second baseline is termed the Hybrid algorithm in \cite{wu2014online}, which combines CUSUM with a MRF model. This algorithm represents one of the most advanced methods for the problem of dynamic ERD in WSN. These baselines are used to demonstrate
the capability and advantage of our model for this problem.

The evaluation metrics used here were borrowed from \cite{wu2011distributed}. They are the miss detection rate (MDR), false alarm rate (FAR) and error rate (ER) defined as follows:
\begin{equation}
\mbox{MDR}=\frac{\sharp\;\mbox{of nodes missing the detection}}{\mbox{Total}\;\sharp\;\mbox{of nodes}}; \nonumber
\end{equation}
\begin{equation}
\mbox{FAR}=\frac{\sharp\;\mbox{of nodes generating false alarms}}{\mbox{Total}\;\sharp\;\mbox{of nodes}}; \nonumber
\end{equation}
\begin{equation}
\mbox{ER}=\frac{\sharp\;\mbox{of nodes making wrong decisions}}{\mbox{Total}\;\sharp\;\mbox{of nodes}}. \nonumber
\end{equation}
According to the above definition, the ER is exactly the summation of MDR and FAR. Note that the above metrics are used to evaluate the accuracy of the algorithms in detecting the event, other than the sensor fault.
\subsection{Simulation Setting}
We simulated a WSN at a square field of 10$\times$10
meters. The number of network nodes was 100. Their locations were uniformly distributed in the field as shown in Fig.\ref{fig:region_expansion}.
In what follows, we use a 2-dimensional (2D) coordinate system to describe relative positions of these nodes, so the left bottom node and the right top node in Fig.\ref{fig:region_expansion} correspond to coordinates $(1,1)$ and $(10,10)$, respectively. For the task of dynamic ERD, we simulated a diffuse event over time by covering a group of sensor nodes. Fig. \ref{fig:region_expansion} shows our experimental setup for data collection, wherein the simulated event originates from an inside region at time $k=1$, and then spreads outwards until $k=5$. In the experiment, two spatiotemporal patterns are assigned to the normal sensor readings of the event region nodes and the non-event region nodes, respectively. Zero-mean Gaussian noises are added to each sensor reading to take into account of the presence of measurement noises. The resulting normal sensing range $[x_{\min},x_{\max}]$ acts as an input to the ERD algorithm.
\begin{figure*}
\centering
\subfloat[k=1]{
\label{region_expansion1}
\begin{minipage}[t]{0.165\textwidth}
\centering
\includegraphics[width=1.4in,height=1in]{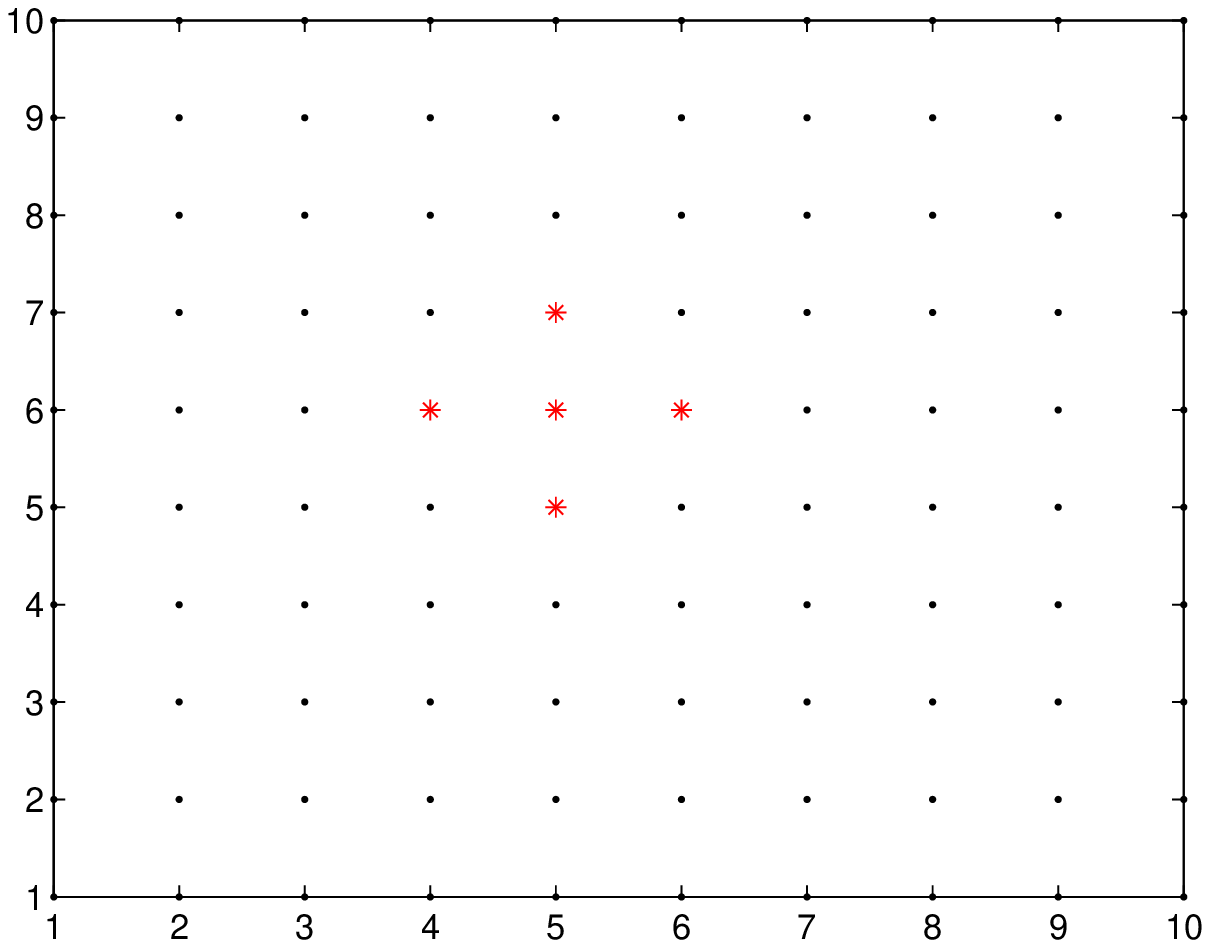}
\end{minipage}
}
\subfloat[k=2]{
\label{region_expansion2}
\begin{minipage}[t]{0.165\textwidth}
\centering
\includegraphics[width=1.4in,height=1in]{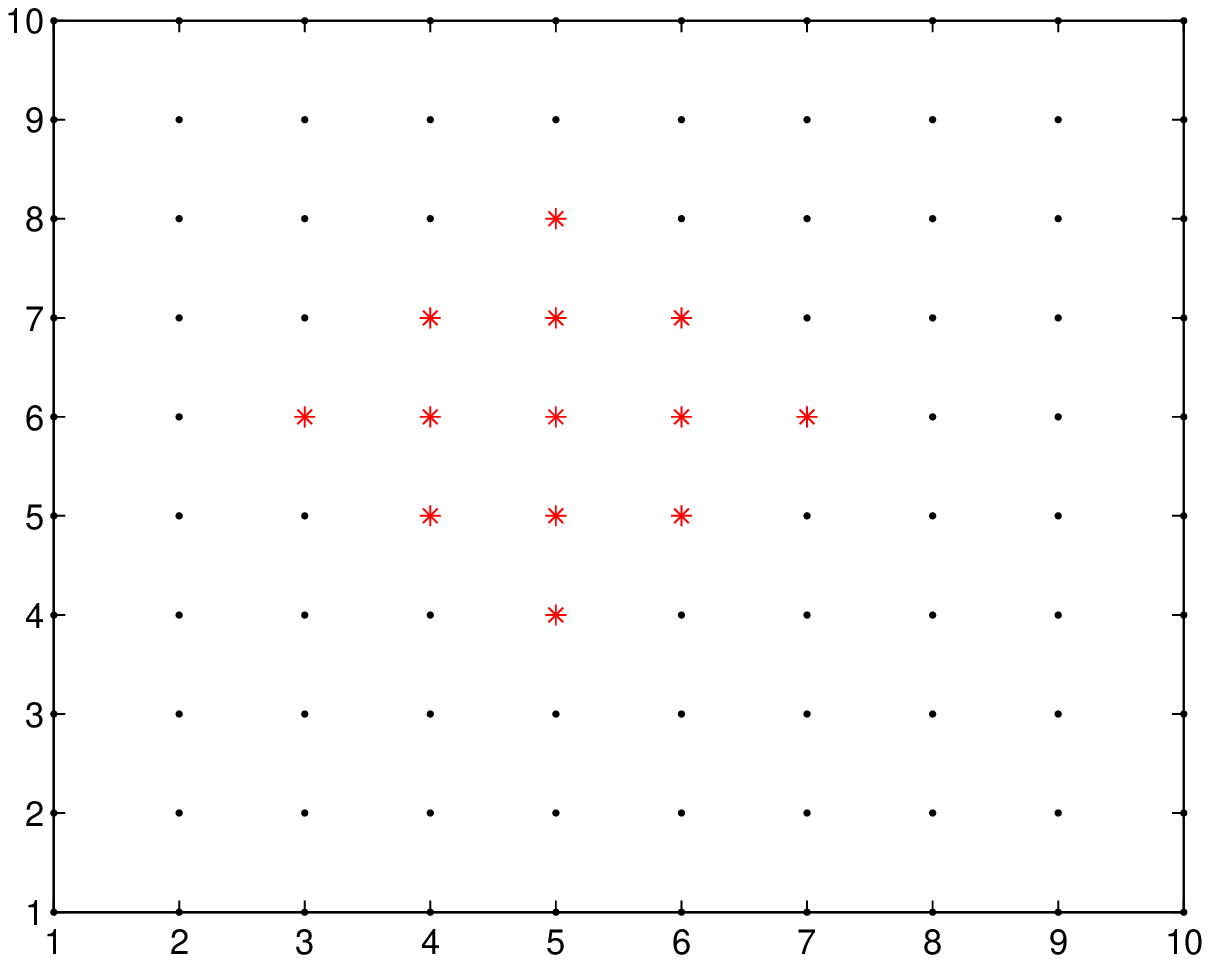}
\end{minipage}
}
\subfloat[k=3]{
\label{region_expansion3}
\begin{minipage}[t]{0.165\textwidth}
\centering
\includegraphics[width=1.4in,height=1in]{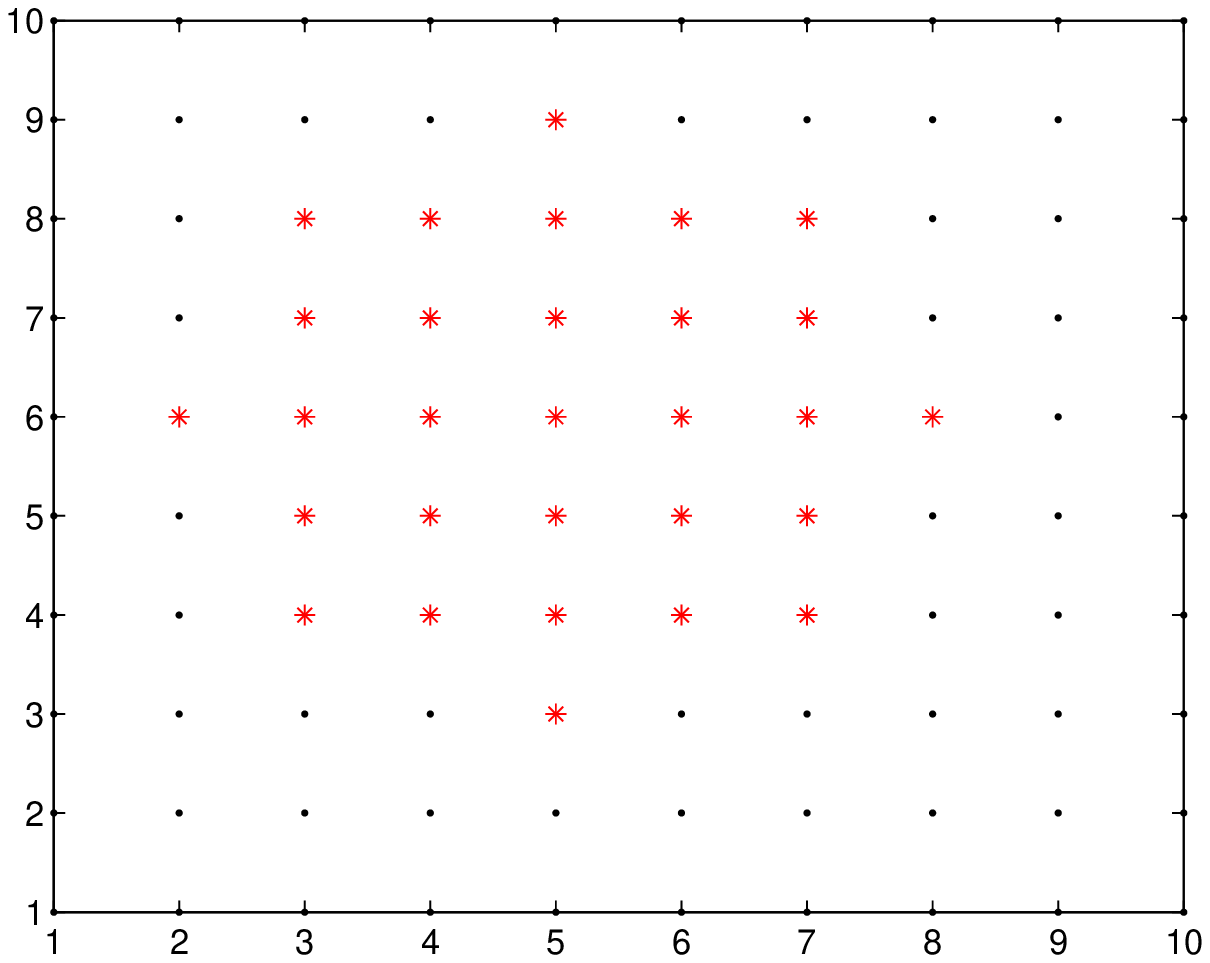}
\end{minipage}
}
\subfloat[k=4]{
\label{region_expansion4}
\begin{minipage}[t]{0.165\textwidth}
\centering
\includegraphics[width=1.4in,height=1in]{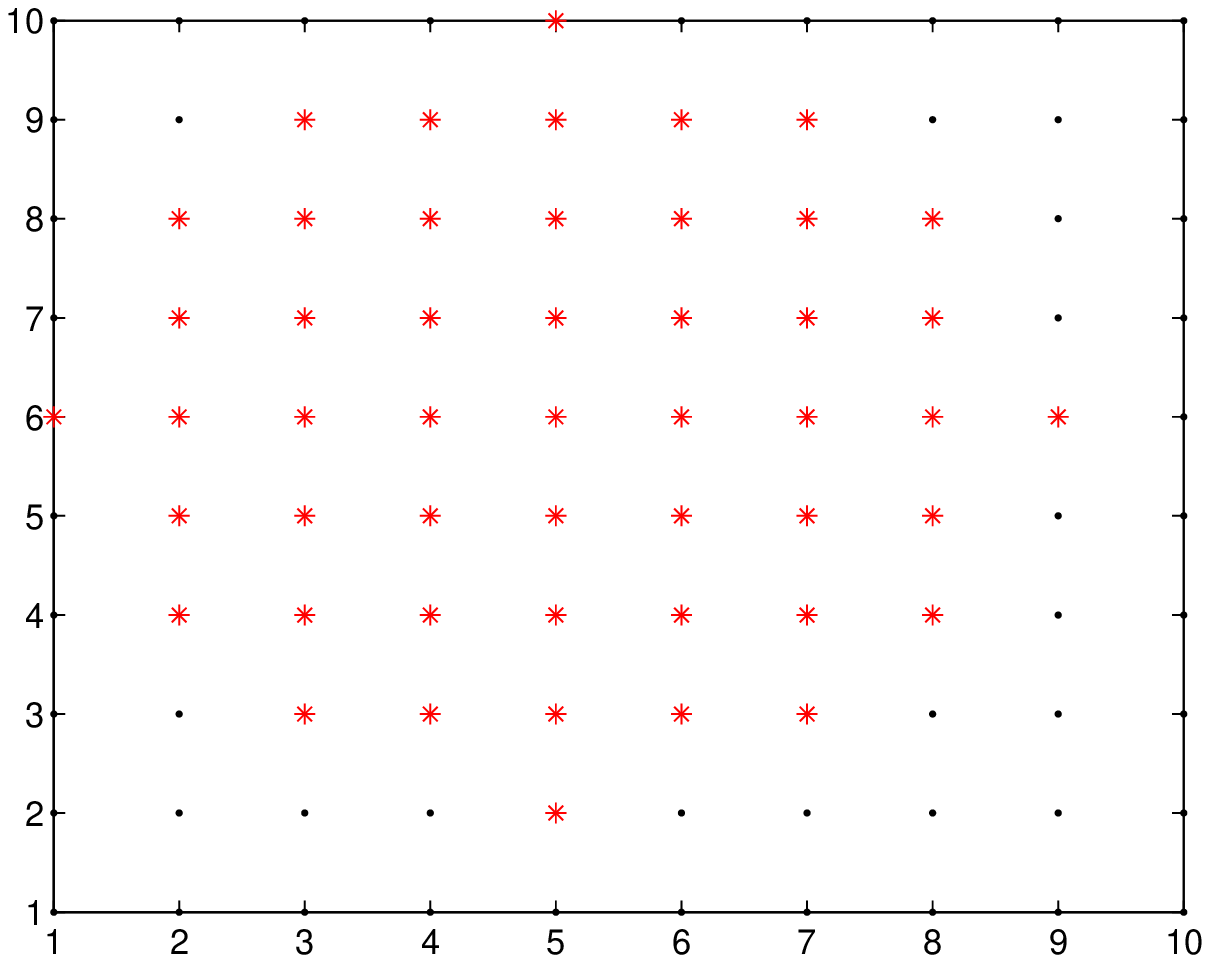}
\end{minipage}
}
\subfloat[k=5]{
\label{region_expansion5}
\begin{minipage}[t]{0.165\textwidth}
\centering
\includegraphics[width=1.4in,height=1in]{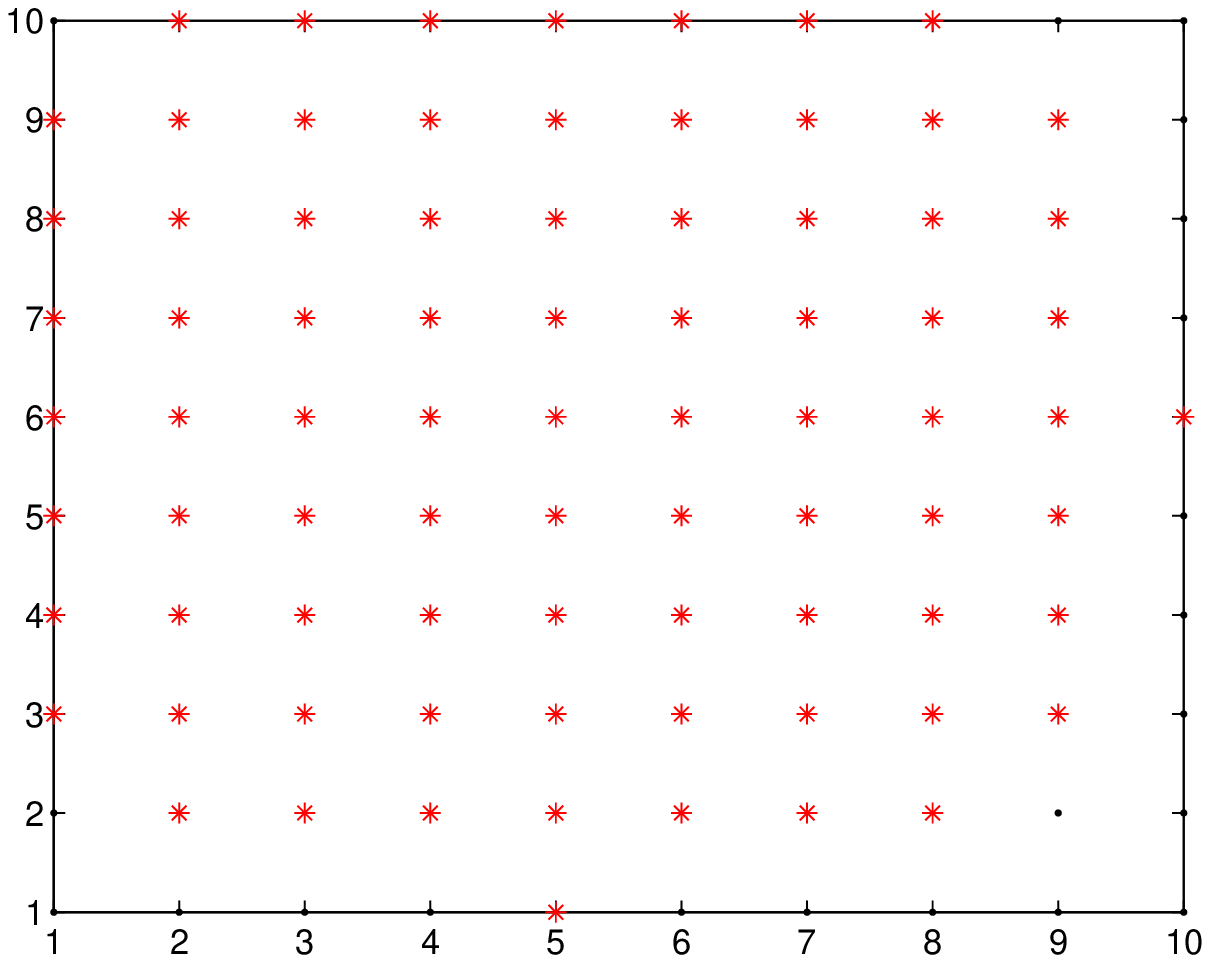}
\end{minipage}
}
\caption{This simulated case of event region diffusing. ``$\ast$" represents nodes covered by the event, ``$\cdot$" represents the others.}\label{fig:region_expansion}
\end{figure*}

The faulty sensor readings are simulated based on the fault models presented in \cite{ganeriwal2008reputation}. Specifically, we considered four different types of faults, termed ``offset fault", ``stuck-at fault", ``variance degradation fault" and ``sleeper attacks", respectively \cite{ganeriwal2008reputation}. Denote $x$ and $x_f$ to be a pair of pre-fault and post-fault senor readings. The ``offset fault" represents the case in which $x_f=x+m$, where $m$ denotes a constant offset. The ``stuck-at fault" represents a sensor getting stuck at a particular value $s$. The ``variance degradation fault" can be represented as $x_f=x+\mathcal{N}(0,q)$. The ``sleeper attacks" represents the scenario, where the node does not transmit any data reading, as if it does not exist in the network. For this scenario, the sensor's \emph{trust} value is estimated based solely on the time-evolution law specified in Eqn.(\ref{transition}). Accordingly we have $\hat{\lambda}_k=\alpha\hat{\lambda}_{k-1}$. We use a parameter $p$ to denote the sensor failure rate, namely the ratio of the number of faulty nodes to that of all nodes in the network under consideration. In the simulation, every node has the same chance to be selected to act as a faulty node and its associated fault type is also specified in a random way.

The setting for the free parameters of the proposed algorithm used in the experiments is presented in Table I.
\begin{table}[H]
\begin {center}
\caption{Parameter setting for the proposed algorithm in the experiment}
\begin{tabular}{c|c|c|c|c|c}
\hline\hline
$N$ & $\alpha$ & $Q$ & $\beta$ & $r$&$\lambda_{thr}$\\
\hline
20 & 0.85& 0.01 &  0.1  & 5 & 0.3\\
\hline\hline
\end{tabular}
\end {center}
\end{table}
\subsection{Results}
First we fixed the failure rate $p$ at $10\%$ and simulated the data set, and ran the CUSUM, the Hybrid algorithm and the proposed algorithm to analyze the same data set. The experimental results at time steps $k=2,3,4$ are presented in Tables II, III and IV, respectively. It shows that the proposed algorithm is more accurate than the other algorithms in terms of different error rate metrics considered here.
\begin{table}[H]
\begin {center}
\caption{Performance metrics at time step $k=2$}
\begin{tabular}{c|c|c|c}
\hline\hline
  & CUSUM & Hybrid & Proposed \\
\hline
MDR & 0.03 & 0.01 & 0\\
\hline
FAR	& 0	& 0.01	& 0\\
\hline
ER & 0.03 &	0.02 & 0\\
\hline\hline
\end{tabular}
\end {center}
\end{table}
\begin{table}[H]
\begin {center}
\caption{Performance metrics at time step $k=3$}
\begin{tabular}{c|c|c|c}
\hline\hline
  & CUSUM & Hybrid & Proposed \\
\hline
MDR & 0.01 & 0.01 & 0\\
\hline
FAR	& 0.02	& 0	& 0.01\\
\hline
ER & 0.03 &	0.01 & 0.01\\
\hline\hline
\end{tabular}
\end {center}
\end{table}
\begin{table}[H]
\begin {center}
\caption{Performance metrics at time step $k=4$}
\begin{tabular}{c|c|c|c}
\hline\hline
  & CUSUM & Hybrid & Proposed \\
\hline
MDR & 0.04 & 0.02 & 0.01\\
\hline
FAR	& 0.02	& 0	& 0\\
\hline
ER & 0.06 &	0.02 & 0.01\\
\hline\hline
\end{tabular}
\end {center}
\end{table}

Next we conducted a Monte Carlo test, in which each algorithm was run 30 times independently and then we recorded the average error rate of each algorithm at each time step. The result is depicted in Fig.\ref{fig:error_t}. We can see that the proposed algorithm performs much better than the others.

Finally we investigated the algorithm performance at different sensor failure rates. We considered a series of failure rates, and for each one, we ran each algorithm 30 times as before and calculate its average error rate. The result at an arbitrarily selected time step $k=3$ is plotted in Fig.\ref{fig:error_p}. It confirms again that the proposed algorithm beats the others in all cases of different senor failure rates.
\begin{figure}[H]
\begin{tabular}{c}
\centerline{\includegraphics[width=3.3in,height=2in]{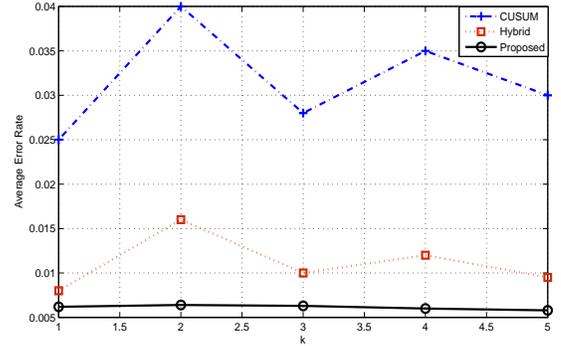}}
\end{tabular}
\caption{Average error rate over the time calculated based on 30 times of independent runs of each algorithm} \label{fig:error_t}
\end{figure}
\begin{figure}[H]
\begin{tabular}{c}
\centerline{\includegraphics[width=3.3in,height=2in]{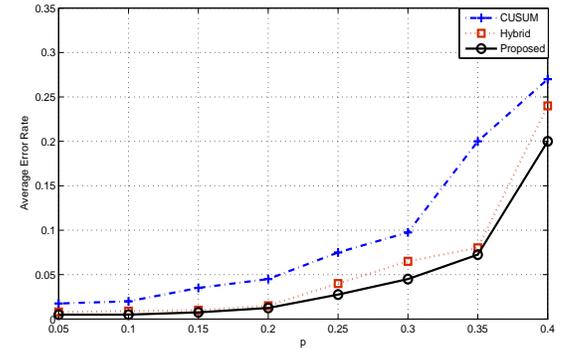}}
\end{tabular}
\caption{Average error rate at different failure rates $p$, calculated based on 30 times of independent runs of each algorithm}\label{fig:error_p}
\end{figure}
\section{Concluding remarks}
This paper presents a model based approach to address the problem of online and fault-tolerant dynamic ERD over WSNs.
Based on the concept of sensor node \emph{trust}, a BTM model is developed here, which links a number of practical issues, such as sensor failures, spatiotemporal correlations of neighbor nodes' sensor readings, temporal evolvement of the event region and the requirement of real-time processing, altogether in the modeling stage.
To the best of our knowledge, the BTM model presented here is the first trust model in the literature that is developed specifically for the problem of fault-tolerant dynamic ERD in WSNs. The proposed ERD algorithm is designed on the basis of the BTM.
The performance of the proposed algorithm is demonstrated by illustrative simulation studies. Experimental results show that our algorithm is more accurate than a traditional baseline method, namely CUSUM, and one of the state-of-the-art methods termed the Hybrid algorithm \cite{wu2014online}, in terms of error rate in detecting the event.
%
%

The success of the proposed solution depends on a basic assumption; that is each sensor node must have a certain amount of computing resources to run the PF algorithm. Since the dimension of the state variable in the BTM model is only 1, the particle size of the PF is just set at 20 in the current solution. So the required amount of computing resources is not great; while, to find more realistic, and especially economical, applications, the PF algorithm used here may still need to be simplified or approximated to save computing resources.

In our current solution, a perfect data transmission between the center node and its corresponding member nodes is assumed. When the above assumption is not satisfied, how to adapt our algorithm accordingly is an issue to be addressed. A related opening problem for applications of our method in a real-life network is how to build up a good enough virtual community for each node, so that a perfect data transmission between this node and its corresponding member nodes can be guaranteed, especially when the network topology is complex, e.g., heterogeneous and/or dynamically evolving.

In our current model, the classification of the monitoring filed is only binary; that is to say, the location of a sensor node either falls within the event region or the non-event region. How to adapt the model and algorithm proposed here to handle more differing classes of monitoring regions is an interesting topic to be addressed in the future.
%
\section*{Acknowledgment}
This work was partly supported by the National Natural Science Foundation (NSF) of China under grant Nos. 61302158 and 61571238, the NSF of Jiangsu Province under grant No. BK20130869, the China postdoctoral Science Foundation under grant Nos. 2015M580455 and 2016T90483, and the Scientific and Technological Support Project (Society) of Jiangsu Province under grant No. BE2016776.
\bibliographystyle{IEEEtran}
\bibliography{mybibfile}
\end{twocolumn}
\end{document}

%% file: title-track.tex
\thispagestyle{empty}
\vspace*{0.6cm}
\begin{center}

\vspace{-.2in}

{\LARGE Online Fault-Tolerant Dynamic Event Region Detection in Sensor Networks via Trust Model
}\\


{\large \vspace{1cm}
\vspace{0.3cm}

Jiejie Wang and Bin Liu\footnote{Corresponding author.
E-mail: {\sf bins@ieee.org}.\\} \vspace{0.5cm}

{$~$School of Computer Science and Technology, \\Nanjing University of Posts and Telecommunications}\\
{$~$Nanjing, Jiangsu, 210023, China\\}

\vspace{1cm}

Manuscript Submitted --- Oct., 7th, 2016\\
\vspace{1cm}
This Manuscript has been accepted by \\2017 IEEE Wireless Communications and Networking Conference (WCNC 2017)\\
} 

\end{center}